**Photophysics of defect-passivated quasi-2D (PEA)$_2$PbBr$_4$ perovskite using an organic small-molecule**

Jafar I. Khan[$,1*], Murali Gedda[$,1], Mingcong Wang[1], Emre Yengel[1], Joshua A. Kreß[2], Yana Vaynzof[2], Thomas D. Anthopoulos[1*] and Frédéric Laquai[1*]

[1]King Abdullah University of Science and Technology (KAUST), KAUST Solar Center (KSC), Physical Sciences and Engineering Division (PSE), Thuwal 23955-6900, Kingdom of Saudi Arabia

[2]Integrated Center for Applied Physics and Photonic Materials, Center for Advancing Electronics Dresden (CFAED), Technical University of Dresden, Nöthnitzer Straße 61, 01187 Dresden, Germany

**Abstract**

2D Ruddlesden–Popper perovskites are promising candidates for energy harvesting applications due to their tunable optical properties and excellent ambient stability. Moreover, they are solution-processable and compatible with upscalable manufacturing via various printing techniques. Unfortunately, such methods often induce large degrees of heterogeneity due to poorly controlled crystallization. Here, we address this issue by blending the well-known 2D perovskite (PEA)$_2$PbBr$_4$ with an organic small-molecule, namely C$_8$-BTBT, employed as an additive with different blending ratios. Using terahertz (THz) absorption and temperature-dependent photoluminescence (PL) spectroscopy techniques we observe that with the C$_8$-BTBT additive the photophysical properties are altered while the perovskite structure in the film remains unaffected. More precisely, the inclusion of trace amounts of C$_8$-BTBT in the hybrid films results in defect passivation at perovskite platelet boundaries and at the surfaces, as indicated by increased carrier lifetimes and substantially increased photoluminescence quantum yields (PLQY). This in turn improves the responsivity of photodetectors using the 2D perovskite as active layer. Our study highlights a straightforward strategy for fabricating high-quality 2D perovskites via large-area processing techniques.

**Introduction**

Low-dimensional perovskites have attracted considerable attention in recent years owing to their versatile (opto)electronic properties and suitability for energy harvesting applications.[1–3] The most extensively studied low-dimensional perovskites are the 2-dimensional Ruddlesden–Popper (2D-RP) perovskite phases of the form $R_2A_{n-1}B_nX_{3n+1}$, where *R* represents bulk hydrophobic alkyl chains or aromatic cations, *A* is a monovalent cation, *B* stands for a divalent metal ion, *X* is a halide anion, and the integer *n* determines the number of stacked semiconducting layers.[4] In these 2D or quasi-2D-RP perovskites, thin sheets of semiconducting inorganic layers are sandwiched between bulky monovalent organic cations *R*, leading to the formation of multiple-quantum-well structures.[5] Due to the substantial quantum and dielectric confinement of charge carriers, these 2D-perovskites (2D PVKs) exhibit large exciton binding energies that can be as large as a few hundreds of meV, unlike their 3D metal halide perovskite counterparts.[6,7] Understanding the fundamental photophysical properties and their dependence on crystallinity and type of the spacer cation of 2D-RP perovskites is essential for further performance improvements of optoelectronic devices by guided material design. Specifically, controlling and combining their intriguing physical properties, such as strong light absorption and efficient photoluminescence (PL), with their enhanced moisture and air stability,[8] could pave the way towards a plethora of applications ranging from photodetectors and light-emitting diodes (LEDs) to electrically-pumped LASER diodes, and energy-harvesting devices.[9–11]

However, despite intense efforts to elucidate the photophysics of 2D PVK, a comprehensive understanding is currently missing as the materials are still in their infancy. The optical properties change with the number of layers as does the quantum confinement. The large dielectric constant difference between the organic and inorganic layers leads to different exciton binding energies.[12–14] In this regard, the photoluminescence quantum yield (PLQY) is an essential property which measures the luminescence efficiency of materials. Generally, the PLQY of 2D PVKs is low due to phase impurities and defects on the film surface originating from the solution processing.[15] To increase the PLQY, various surface modification protocols have been explored; for instance, an impressive PLQY of >80% for $(BA)_2(MA)_4Pb_5Br_{16}$ thin films at 515 nm was achieved through the introduction of the electron donor 9,9-spirobifluoren-2-yl-diphenylphosphine oxide (SPPO1) as well as trioctylphosphine oxide (TOPO) molecules.[16,17] Surprisingly, not much is known about the properties of *n*=1 layered PVK. In such layered PVK systems, the exciton binding energy is large and decreases with increasing *n*, allowing for finetuning of the optoelectronic properties toward various applications.[18,19] This intriguing property results from the quantum confinement that occurs due to the formation of multiple-quantum-well structures within the PVK.[1] Quantum confinement depends on the thickness of the quantum well, the value of $n$, and the barrier, the combination of which can significantly affect the bandgap. It has been predicted and experimentally confirmed that the band gap and exciton binding energy both decrease as $n$ increases. Despite the significant body of experimental evidence, however, a clear classification of excitons in 2D-RP perovskites is not yet available. To this end, the model of Frenkel excitons is often used to explain their properties,[20–22] while the actual radii of the excitons are comparable to the size of the unit cell. This yields a strong exciton luminescence and absorption even at room temperature with short exciton lifetimes typically in the range of picoseconds.[5,20,23]

Here, we study the photophysical properties of the 2D-RP perovskite $PEA_2PbBr_4$ with $n = 1$. Control over morphology and crystallinity was achieved through the use of an organic semiconducting small-molecule, $C_8$-BTBT (2,7-dioctyl[1]benzothieno[3,2-*b*][1]benzothiophene), used as an additive in the 2D-RP perovskite (Figure 1a and b). We first explored the impact of $C_8$-BTBT on the photophysical properties of the resulting hybrid blends. Time-resolved spectroscopy was used as a diagnostics tool to monitor the exciton characteristics in blends with varying compositions. Time-resolved photoluminescence (TR-PL) measurements revealed significantly prolonged PL lifetimes for 25% (volume ratio, v/v%) blend systems as compared to that of pure PVK films. The opposite is observed for the PL lifetime of 50% samples, which exhibit substantially shortened PL lifetime when compared with the pristine films. The prolonged excitonic PL lifetime observed for the 25% blend systems is assigned to reduced trap density and minimization of non-radiative recombination at the surface and in the bulk of the PVK layers. Moreover, the PLQY is found to increase from 10% to 25% upon $C_8$-BTBT addition. Terahertz (THz) measurements revealed the phonon modes of the 2D-PVK and the C8-BTBT, respectively. We found that the presence of $C_8$-BTBT had no impact on the perovskite lattice, although it shifted the relative peak position of the phonon mode; an effect attributed to the increased size of the platelets-like 2D-perovskite domains. From transient absorption (TA) measurements, we observed a pronounced spectral shift of the photoinduced absorption region with time, which we assigned to bandgap renormalization. Furthermore, the carrier dynamics were significantly slowed upon the addition of $C_8$-BTBT, which is consistent with the TR-PL transients. Finally, we establish a correlation between photophysical and optoelectronic properties by realizing photodetectors using different blend formulations. The device dynamics indicated that introducing $C_8$-BTBT at an optimal concentration into the PVK helps boosting the responsivity of the device photocurrent.

**Results and discussion**

   1. **Material characterization**

The fabrication and morphological characterization of the representative materials are described in this section. The various blend formulations are detailed in the experimental section. In brief, blends were prepared by mixing the individual precursor formulations of (PEA)$_2$PbBr$_4$ and C$_8$-BTBT in three different perovskite/small-molecule volume ratios (*v/v*): 90/10 *v/v* (10% blend), 75/25 *v/v* (25% blend), and 50/50 *v/v* (50% blend). We adopted the spin-coating technique to process the thin films of all formulations at room temperature inside a nitrogen-filled glovebox followed by thermal annealing, as detailed in the experimental section. Using X-ray diffraction (XRD) measurements, we investigated the crystallinity of the emergent layers, the corresponding diffractograms of pristine (PEA)$_2$PbBr$_4$ and the three representative blend thin films are presented in Figure 1c. Notably, the evenly spaced sharp diffraction peaks of (002) up to (00 10) confirm the formation of the single-layered (*n*=1) stacks of 2D-RP perovskite phase with preferential out-of-plane orientation in all samples.[24] Further analysis of the diffractograms show that the 25% blend layers exhibit more intense diffraction and reduced peak widths, emphasizing the key role of C$_8$-BTBT in mediating the crystallization and growth of the (PEA)$_2$PbBr$_4$ domains without altering the 2D-RPP phase. The peak at (002) was analyzed using the Debye-Scherrer method and reveals a significant increase in the crystallite size from ≈75 nm for neat (PEA)$_2$PbBr$_4$, to ≈123 nm for the 25% blend (Table S1, *Supporting Information* (SI)). The significant augmentation in the crystalline domains is expected to benefit to charge carrier transport due to fewer grain boundaries. To check the morphology changes of different mixing ratios, we further conducted polarized-light optical microscopy (POM), images of the formed layers are shown in Figure 1d-g. All layers appear heterogeneous and their formulation undergoes a clear and abrupt change during deposition, especially, the 25% blend film yields layers composed of large platelet-like domains.

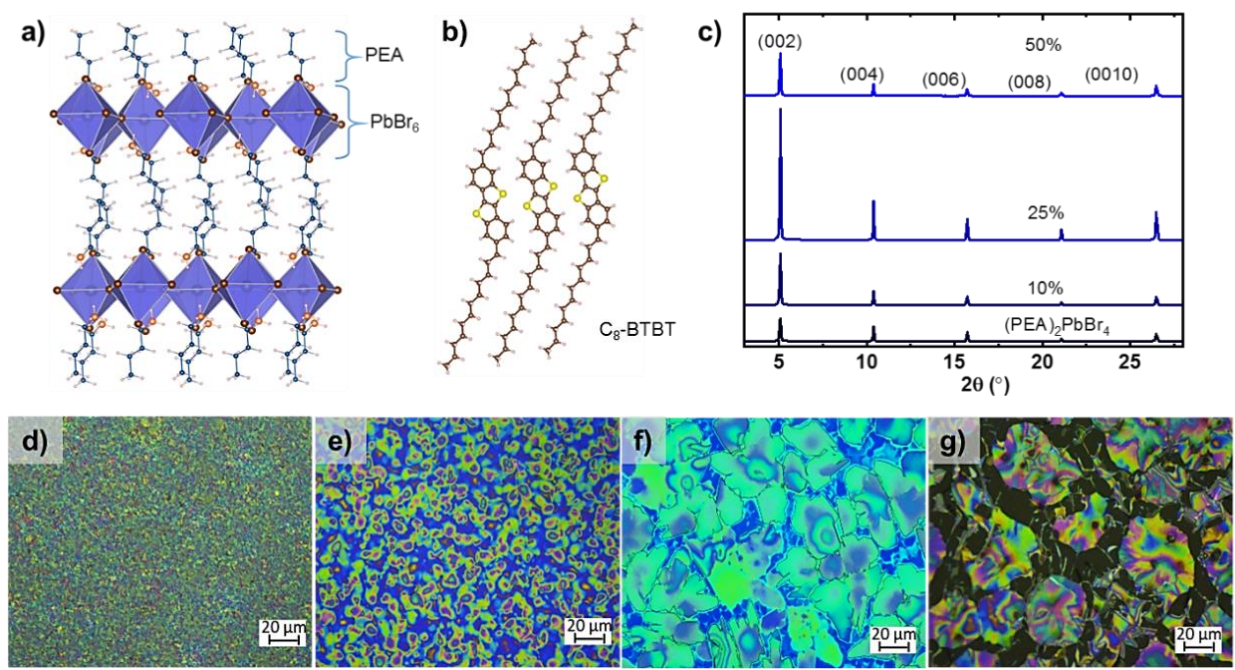

**Figure 1**. (a) Molecular structures of the 2D perovskite $(PEA)_2PbBr_4$ and (b) organic C8-BTBTall-molecule $C_8$-BTBT. (c) Out-of-plane XRD patterns were measured for $(PEA)_2PbBr_4$ and all three blend films. Optical micrographs of (d) pristine $(PEA)_2PbBr_4$, (e) 10% blend, (f) 25% blend, and (g) 50% blend perovskite thin-films.

To examine the composition of crystalline domains of the blend films, X-ray photoemission spectroscopy (XPS) depth profiling experiments were performed on one of the blend films, namely, the 25% blend (Figure S1). We confirmed from the measurements and also from our previous work that the perovskite domains are passivated by a thin $C_8$-BTBT layer during the film formation process (Figure S1a).[25] This surface layer is estimated to be approximately 4 to 5 nm thick and is likely a result of the lower surface energy of $C_8$-BTBT than that of the perovskite film. As shown in Figure S1b, SI, up to ~4 nm etching depth (Figure S1a, zone-1, SI), the elemental profile is dominated by the $C_8$-BTBT. Beyond, the perovskite contribution is more pronounced (Figure S1a, zone-2, SI) until the ITO substrate is reached (Figure S1a, zone-3). Material percentages extracted from the elemental profiles (Figure S1b) showed that below the surface layer, $C_8$-BTBT contributes only ~10% to the overall composition. Since XRD measurements confirmed that no changes to the 2D perovskite crystal structure occur upon blending with the $C_8$-BTBT (Figure 1c), the contribution of $C_8$-BTBT throughout the depth profile is likely to originate from the organic molecule coating the walls of perovskite grains and the substrate (Figure S1b, inset).

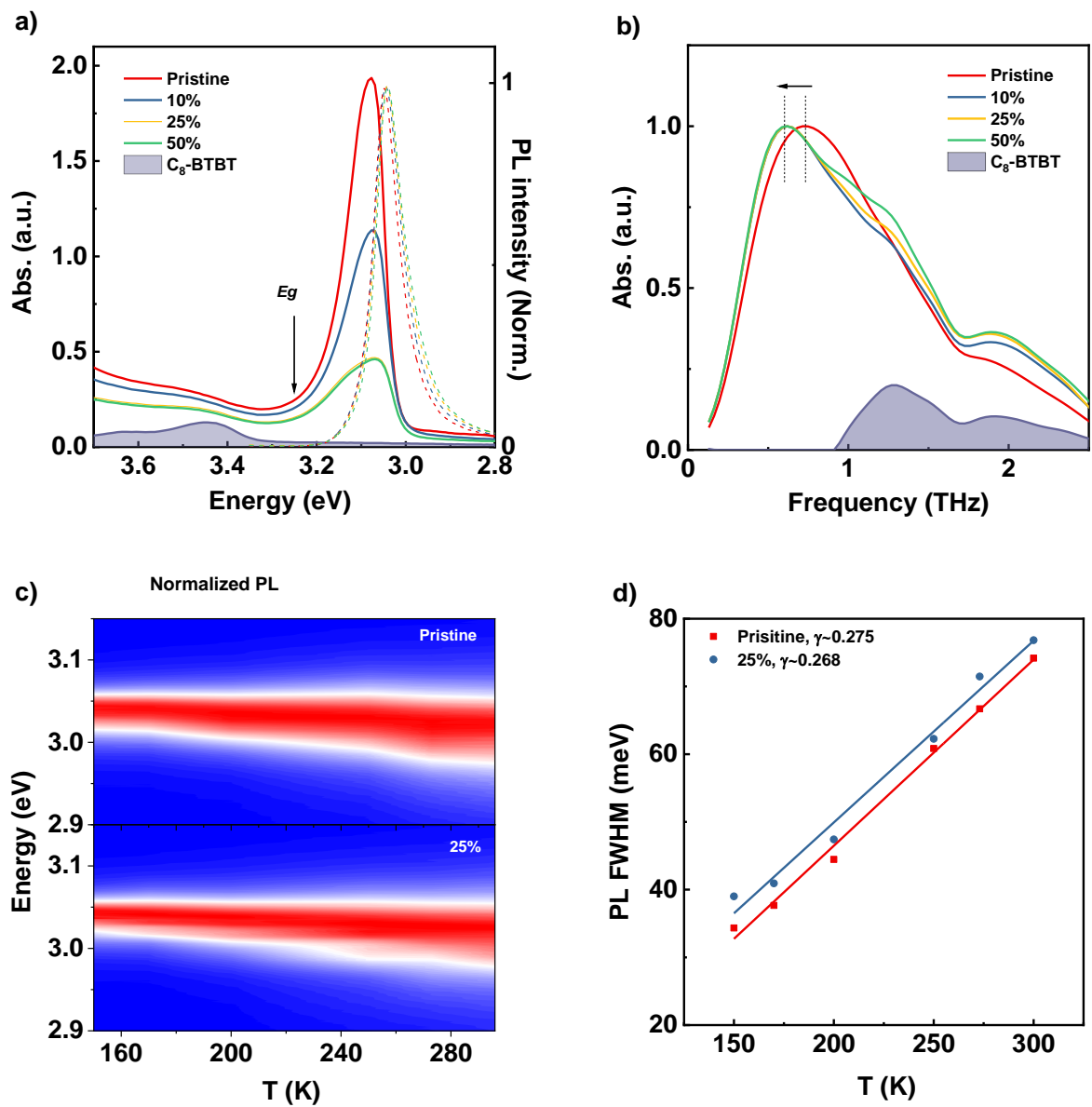

**Figure 2.** Absorption and PL spectra of the pristine 2D-perovskite film and the respective 10, 25 and 50% blends. In (a) absorption spectra (solid lines) and steady-state PL spectra (dashed lines) are shown. The bandgap energy ($E_g$) is extracted from the Tauc plot and is indicated by the black arrow; in fact, all samples have similar $E_g$. In (b) the THz spectra of pristine 2D-perovskite (top panel) and of the 25% blend (bottom panel) at different temperatures are shown, (c) shows the temperature-dependence of the PL peak of the spectra, and (d) shows a linear fit to the line-width of the PL, in which the slope $\gamma$ represents the electron-lattice coupling constant.

## 2. Steady-state optical characterization

We further performed optical characterization by UV-vis spectrophotometry on neat perovskite films, C$_8$-BTBT films, and 10, 25 and 50% blends, results displayed in Figure 2a (left axis). The pristine 2D-perovskite thin-film exhibited a strong exciton resonance at ~3.07 eV in the absorption spectrum ascribed to the enhanced attractive potential of electrons and holes due to the quantum confinement effect.[26] The bandgap energy ($E_g$) was determined from a Tauc plot yielding a value of ~3.27 eV and an exciton binding energy of $E_b$~200 meV, which is well above the thermal energy at room temperature (~25 meV). This indicates that the photophysics in this system is mainly determined by excitons, which is corroborated by the steady-state PL measurements. Specifically, Figure 2a (right axis) shows that the PL spectrum is close to the exciton resonance and far from $E_g$, indicating that the emission is predominantly dictated by the radiative recombination of excitons rather than free carriers.[27,28] All samples have similar $E_g$, $E_b$, and PL spectral distributions (right axis in Figure 2a), indicating the addition of C$_8$-BTBT hardly affects the 2D-perovskite lattice.

The absorption spectra of the blends exhibit a negligible contribution from C$_8$-BTBT due to its intrinsically low absorption coefficient. The absorption peak of the C$_8$-BTBT is located around 3.4 eV as observed in Figure 2a. From Figure S1, SI, we observe that C$_8$-BTBT is present, primarily near the film surface, forming an ultra-thin layer. The reduced exciton resonance upon the addition of C$_8$-BTBT is likely due to the reduction of 2D-perovskite's volume ratio. To further confirm the existence of C$_8$-BTBT, THz absorption spectra were collected for all samples (Figure 2b). The characteristic phonon modes of pristine 2D-perovskite are located at ~0.73 THz and ~1.8 THz. When the volume ratio of C$_8$-BTBT increases, the 0.73 THz peak shifts slightly towards lower energies. We further noted the absorption enhancement at both ~1.25 THz and ~1.92 THz in the blend samples, which coincides with the characteristic phonon modes of C$_8$-BTBT. This verifies that the presence of C$_8$-BTBT hardly affects the 2D-perovskite lattice. Furthermore, it was inferred that the C$_8$-BTBT encapsulation is not controlling the electron-lattice coupling of the 2D-perovskite. Figure 2c shows the temperature dependence of the PL with peak energy plotted as a function of temperature for the pristine perovskite and the 25% blend. To conclude, the incorporation of C$_8$-BTBT does not alter the temperature dependence of the PL as no variation of the peak position was observed in either sample. More precisely, Figure 2d displays fits of the line-width $\Gamma$ of the PL of both representative samples by $\Gamma=\Gamma_0+\gamma T$, where $\Gamma_0$ represents the PL line-width at absolute zero temperature, and $\gamma$ is the electron-lattice coupling constant.[29] On the basis of these results, we inferred that the electron-lattice coupling constants in both pristine and 25% blend samples are similar, indicating $C_8$-BTBT hardly influences the interaction between the charge carriers and phonons.

### 3. Exciton dynamics

The effect of $C_8$-BTBT encapsulation on the crystalline domains as well as on the exciton dynamics was investigated through hyperspectral PL imaging. The acquired images of the neat perovskite and the 10, 25 and 50% blends are presented in Figure 3. Clearly, the crystalline domain size increased upon successive addition of $C_8$-BTBT as shown in Figure 3a-d. The results also support the crystallite size calculations summarized in Table S1. When the PL spectra of each pixel of the hyperspectral images (1040 ×1392 pixels) are analyzed further, the histogram of the PL peak positions' maxima remained at 417 nm (Figure 3e), indicating the uniformity of the crystals in the domain. Also, in Figure 3e, the increased histogram peak at 418 nm observed at higher blend ratio supports the slight shift of the PL spectra as detected by the steady-state PL measurements in Figure 2a.

To study the exciton dynamics, we performed transient absorption (TA) and time-resolved photoluminescence (TR-PL) measurements following photoexcitation at 380 nm of the different samples and monitored at the peak position of 415 nm (Figure 3f-g). The PL spectra (Figure 3f) show narrow bandwidth with a full-width at half-maximum (FWHM) around 12-14 nm. Upon blending, we observed that the 10% sample showed a slightly longer PL lifetime than the pristine sample, and the 25% sample exhibited the longest lifetime. In contrast, the PL lifetime was markedly reduced for the 50% sample compared to the others. The weighted-average lifetimes ($\tau_{avg}$) were extracted by fitting the kinetics to a two-exponential function, and the values are 269, 341 and 144 ps for the respective blends (10, 25 and 50%), whereas the PL lifetime was 181 ps for the pristine sample, all parameters provided in Table S2, SI. We assign the prolonged carrier lifetime to a reduction in the trap density and the trap-associated non-radiative recombination. This implies that the non-radiative recombination pathways are reduced upon adding $C_8$-BTBT up to 25%; however, the 50% sample exhibited a shorter PL lifetime indicating increased non-radiative recombination, which is clear from the initial rapid decay in the first 200 ps. Thus, a trade-off exists in the addition of $C_8$-BTBT. Excessive addition instigates to the formation of trap states and increases non-radiative recombination. In fact, it is apparent that in the first 100 ps the decay is significantly faster in the 50% sample, while after 600 ps the decay is similar to that in the 25% sample. This slower decay was assigned to electron-hole recombination. The narrow bandwidth facilitates applications of these materials in light-emitting devices. In particular, the improved charge transport enabled by enhanced crystallinity and reduction of non-radiative recombination in the 25% sample is expected to lead to more efficient light conversion, which is shown below for a photodetector using the 25% blend.

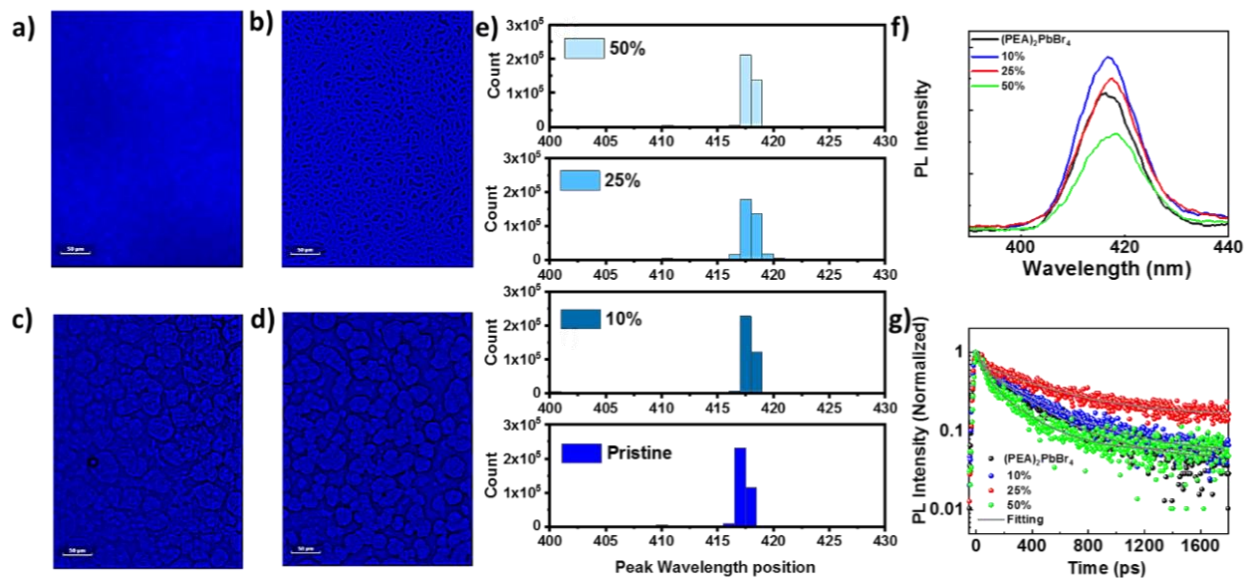

**Figure 3:** Hyperspectral images of a) pristine, b) 10 %, c) 25% and d) 50% blend perovskite films taken at 417 nm. The scale bar is 50 µm. g) Histogram of the peak wavelength positions of the PL spectra of each pixel shown in Figure a-d. f) TR-PL time-integrated spectra of the representative samples taken after optical excitation at 380 nm using a laser fluence of 400 nJ/cm$^2$. g) The transients tracked at the PL peak position of 415 nm and the respective fits (solid grey lines).

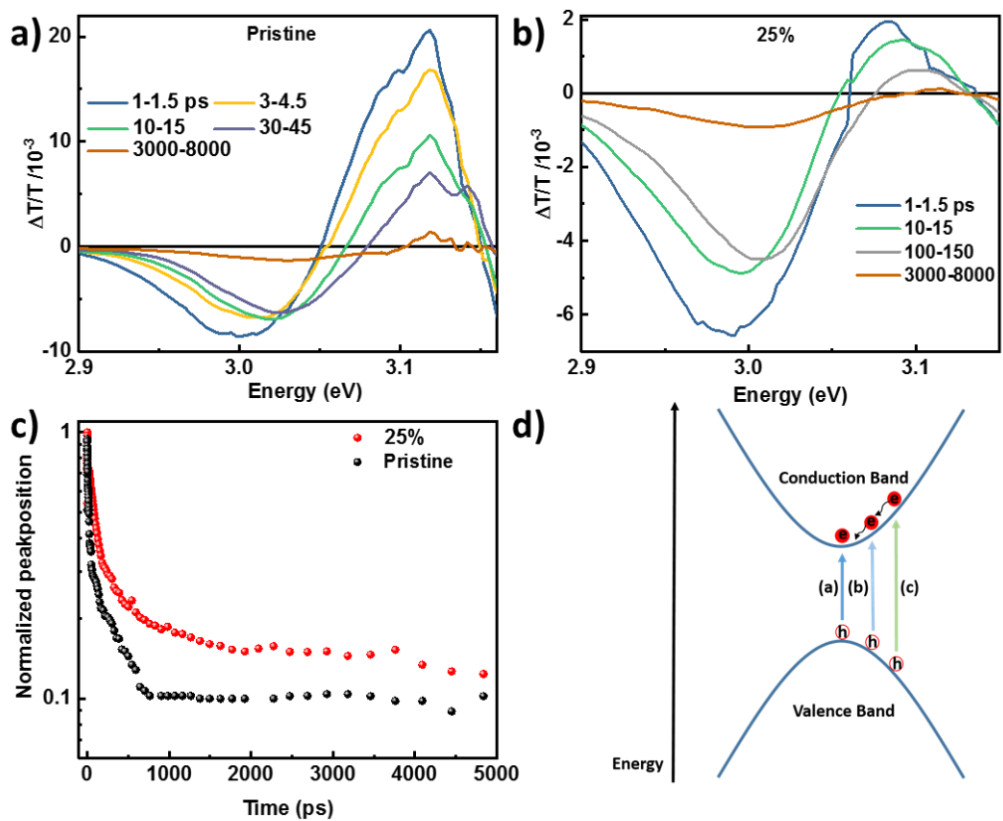

**Figure 4:** a) TA spectra of the pristine 2D perovskite and b) the 25% blend at different delay times after photoexcitation at 380 nm. c) TA kinetics for the pristine and 25% sample, showing the shift of the peak position versus delay time. d) Schematic band model explaining the origin of the temporal spectral shift due to excitation into higher-lying excitonic states.

We performed *ps-ns* transient absorption (TA) spectroscopy up to 8 ns delay time to evaluate the excited state dynamics. All measurements were conducted after optical excitation at 380 nm using a moderate laser fluence of 10 µJ/cm$^2$. The TA results of both the pristine and 25% blend thin films (Figure 4a-b) at different delay times reveal a similar spectral shift with time. The TA spectra of all four samples are presented in Figure S2. We identify the following spectral regions of interest, namely, the ground state bleach (GSB) region in the range of 3.05-3.15 eV, the photoinduced absorption (PA) in the range of 2.9-3.04 eV, the bulk perovskite bleach with a peak at 2.8 eV, and a broad spectral range covering 1.4 -2.7 eV. However, the bulk perovskite photobleaching is not detectable due to the weak signal intensity of the 25 and 50% samples (Figure S2, SI) as the PA contribution is substantial in these systems. The difference in PA of the pristine and 10% sample is significant as a spectral shift occurs over time. The shift is most pronounced in the neat 2D perovskite when compared to the blends. We assigned the spectral shift to bandgap renormalization, as reported earlier.[30] Additionally, we attribute the apparent shift to an effective decrease of the bandgap renormalization stemming from the exciton recombination. This recombination includes both exciton and electron-hole recombination. Noteworthy, the shift occurs faster for the pristine sample as seen from the time-integrated spectra which shift already at 3 ps. However, after 30-45 ps the spectral position remains constant over the entire 8 ns time window. In contrast, in the 25% sample, the onset of the shift is delayed to 10 ps, and concluded after 150 ps with no further shift occurring in the 8 ns time window. Additionally, the energy range of the shift is smaller in the 25% blend sample, whereas the pristine sample exhibits a larger shift across a shorter time range. For further evaluation, we plotted the energy shift versus time delay and observed that the signal of the blend decays slower compared to that of the pristine, as seen in Figure 4c. In fact, we observe the faster shift occurs in the pristine sample, and further that the non-radiative recombination is reduced in the 25% blend sample as reflected by the prolonged dynamics, which is consistent with the TR-PL kinetics. This spectral shift with time in the PA range is assigned to different excitonic transitions and most probably higher-lying excitonic states and their relaxation pathways as illustrated in Figure 4d. More precisely, upon exciton generation, the electrons are promoted to higher-lying states above the conduction band minimum (CBM). Subsequently, they undergo relaxation and reach the lower-lying (band edge) states in the vicinity of the CBM. Consequently, we attribute the temporal spectral shift to the aforementioned processes.

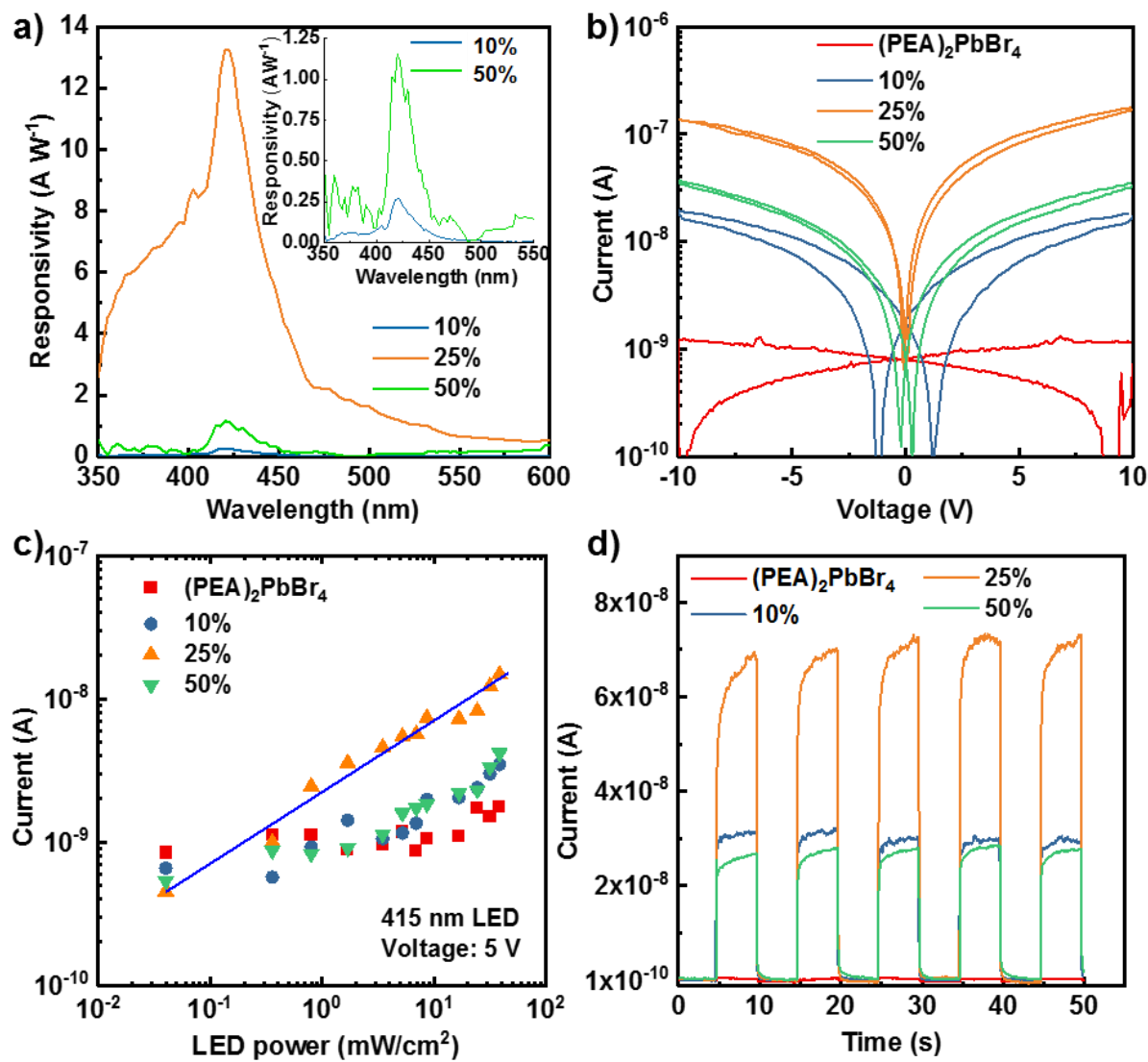

**Figure 5:** a) Responsivity spectra of all three perovskite blend photodetector at 5 V forward biases, b) Current-voltage characteristics of all four photodetectors under constant illumination at 415 nm LED with 4.3 mW/cm$^2$ power. In c) Intensity-dependent photocurrent measurements of the pristine and blend photodetectors at different illumination power densities of 415 nm wavelength LED, and d) Current–time measurements of all four photodetectors at 5 V under 415 nm LED illumination (4.3 mW/cm$^2$ power).

Photodetector devices with a co-planar geometry were fabricated from three different blends. Figure 5a shows the detectors' spectral photoresponses in the range of 350 to 470 nm. A pronounced increase of the spectral photoresponse around 470 nm is observed with a peak value at 425 nm. We attribute this to the generation of electron-hole pairs upon excitation with photons of energy larger than the perovskite bandgap (~3.2 eV).[31] Remarkably, devices based on the 25% blend yielded increased photoresponsivity (R) with increasing illumination energy, with a peak value of 13 mA W$^{-1}$ measured at 425 nm under 5 V. The enhanced R is ascribed to the longer carrier lifetime and lower charge trap density in single-crystalline structures as confirmed by the transient measurements, which is also consistent with the higher PLQY of this blend. A appealing attribute of the photodetector based on the 25% blend layer is its negligible hysteresis, which indicates trap passivation by C$_8$-BTBT, as well as the higher quality of the perovskite phase present in optimized blends. In contrast, neat, 10%, and 50% blend devices exhibit hysteresis as shown in Figure 5b, most likely due to the presence of mobile ionic defects that are known to exist in solution-processed PVK.[32] Upon illumination with photons from a 415 nm LED, linear and symmetric photocurrent versus voltage (I–V) curves were observed as shown in Figure 5c, indicating an ohmic contact between the perovskite phase and the electrodes. Upon illumination with an incident power density in the range of 0.04 to 38.44 mW/cm$^2$, the photodetector based on the 25% blend yielded the most linear response of all devices with a photocurrent modulation from $5 \times 10^{-10}$ A to $3 \times 10^{-8}$ A (Figure 5c). We attribute this to efficient trap passivation achieved at this blend formulation. Figure 5d shows the photocurrent response to photoexcitation pulses at 10 s intervals. All the photodetectors' dynamic photoresponses are very stable, indicating that the devices exhibit a remarkable photoswitching behavior. Devices based on the 25% blend layers exhibit superior photoresponse with a maximum ON-OFF ratio in the order of $10^3$.

**Conclusion**

We reported on the impact of a small organic molecule, namely C$_8$-BTBT, used as additive to passivate traps in a 2D perovskite and demonstrated the use of such blends in photodetectors. The effect of C$_8$-BTBT on the 2D perovskite phase was studied by transient optical measurements. The measurements showed that the carrier dynamics were slowed in 25% blends as compared to neat perovskite, 10%, and 50% blends. Optimal blends displayed improved morphology, reduced non-radiative recombination, and higher PLQY due to the passivation of defects in the bulk and surface of the films. Optimized PVK-C$_8$-BTBT layers were successfully used to realize co-planar photodetectors with enhanced photoresponse and reduced hysteresis.

**Experimental part**

**Sample protocol**: The (PEA)$_2$PbBr$_4$ precursor solution of 0.2 M concentration was prepared by dissolving 1:2 molar ratio of PbBr$_2$ and C$_6$H$_5$C$_2$H$_4$NH$_3$Br (PEABr) in dimethylformamide (DMF) while heated at 75 °C and under dynamic mixing for 5 h. The C$_8$-BTBT solution of 0.02 M was prepared separately using chlorobenzene (CB) as the solvent. The blend precursor formulations were prepared with three different perovskite/organic volume ratios (vol%) of: 90/10 (Blend-I), 75/25 (Blend-II) and 50/50 (Blend-III). The precursor solutions are clear and transparent in color and did not undergo precipitation even after two weeks.

**UV-vis and Photoluminescence (PL) spectroscopy**: Absorption spectra were obtained using an Agilent Cary 5000 UV–vis–NIR spectrometer. Steady-state PL measurements were carried out using a Horiba Aramis Raman system (λ= 325 - 1064 nm, P≤300 mW, LASER) in ambient conditions.

**Hyperspectral Imaging:** Hyperspectral PL images were recorded with an IMA hyper-spectral microscope from Photon ETC. A 405 nm CW laser was used as an excitation source and spectral images were collected from 405 nm to 450 nm on an area of 330 μm × 442 μm with 10 seconds integration time for 1 nm steps.

**X-ray diffraction spectroscopy**: Bruker D8 ADVANCE diffractometer with a monochromatic Cu - Kα radiation beam with wavelength λ = 0.154 nm was used for X-ray diffraction measurements. The thin films were studied in powder diffraction mode using θ–2θ scan configuration to detect crystallites with a reciprocal lattice vector perpendicular to the film surface.

**Photodetector fabrication and characterization**: Photodiode fabricated in top contact geometry on heavily doped Si (Si$^{++}$) wafers with 300 nm thermally grown SiO$_2$, acting as substrate. Before thin-film deposition, the Si$^{++}$/SiO$_2$ wafers were thoroughly cleaned using a series of ultra-sonication baths of de-ionized (DI) water, acetone and 2-propanol for 10 min each, followed by a UV-ozone treatment step for 15 min. The pre-prepared formulations were spun on Si$^{++}$/SiO$_2$ substrates with a 2000 rpm spinning rate for 30 s, followed by thermal annealing at 70 °C for 5 min in the dry nitrogen atmosphere. Photodetector fabrication was completed by the thermal deposition of a top Au electrode of 50 nm thick under a high vacuum (~1 × 10$^{-6}$ mbar). The resulting devices' channel width (W) and length (L) were 1000 μm and 40 μm, respectively. Current-voltage characterization of the device at room temperature was carried out in a nitrogen glove box using a precision source/measure unit, B2912A (Keysight Technologies). The photodetector's responsivity was tested using a Oriel monochromator with a calibrated Si photodetector as a reference. Light Intensity dependent measurements were performed using a Thorlabs M415L4 controlled with Throlabs DC2200 LED Driver. A home-built MATLAB script has been employed to probe various characteristics of the photodetectors.

**Transient absorption (TA) spectroscopy**: TA spectroscopy was carried out using a home-built pump-probe setup. The configuration of the setup was applied for the short delay, namely 100 fs to 8 ns experiments. The fundamental 800 nm output of titanium:sapphire amplifier (Coherent LEGEND DUO, 4.5 mJ, 3 kHz, 100 fs) was split into three beams (2 mJ, 1 mJ, and 1.5 mJ). The first two components were used separately to pump two optical parametric amplifiers (OPA) (Light Conversion TOPAS Prime). TOPAS 1 generates tunable pump pulses, while TOPAS 2 generates signal (1300 nm) and idler (2000 nm) only. For short delay TA measurements, TOPAS 1 was used for producing pump pulses while the probe pathway length to the sample was kept constant at approximately 5 meters between the output of the TOPAS 1 and the sample. The pump pathway length was varied between 5.12 and 2.6 m with a broadband retroreflector mounted on automated mechanical delay stage (Newport linear stage IMS600CCHA controlled by a Newport XPS motion controller), thereby generating delays between pump and probe from -400 ps to 8 ns. We used an 800 nm beam focused onto a sapphire crystal to measure TA in the completely visible range, thereby generating a white-light supercontinuum from 500 to 1300 nm.

Pump and probe beams were focused on the sample, which was kept under a dynamic vacuum of <$10^{-5}$ mbar. The tranC8-BTBTitted fraction of the white light was guided to a custom-made priC8-BTBT spectrograph (Entwicklungsbüro Stresing) where a priC8-BTBT dispersed it onto a 512 pixel NMOS linear image sensor (HAMAMATSU S8381-512). The probe pulse repetition rate was 3 kHz, while the excitation pulses were mechanically chopped to 1.5 kHz (100 fs to 8 ns delays), while the detector array was read out at 3 kHz. Adjacent diode readings corresponding to the tranC8-BTBTission of the sample after excitation and in the absence of an excitation pulse were used to calculate ΔT/T. Measurements were averaged over several thousand shots to obtain a good signal-to-noise ratio. The chirp induced by the tranmissive optics was corrected with a home-built Matlab code by revaluating for each wavelength the delay at which pump and probe are simultaneously arriving on the sample as the time of the signal amplitude.

**Time-domain terahertz (td-THz) spectroscopy:** Our td-THz setup uses the same Ti:sapphire amplifier as the TA setup. The THz emitter and detector are two 1 mm thick <110> oriented zinc telluride (ZnTe) crystals. All the THz related optics were placed in a closed chamber, which was continuously purged with pure nitrogen gas.

**Time-resolved photoluminescence (TR-PL):** The measurements were performed by exciting the samples with the Chameleon Ultra laser (Coherent) at 760 nm with a pulse width of 140 femtoseconds (fs) and a repetition rate of 80 MHz. The fundamental of the Chameleon is pumping a Radiantis OPO, from which the generated second harmonic 380 nm is used to excite the samples. The repetition rate of the fs pulses was adjusted by a pulse picker (APE Pulse Select), and typically the pulse energies were in the range of 400 nJcm$^{-2}$). The PL of the samples was collected by an optical telescope (consisting of two plano-convex lenses), and focused on the slit of a spectrograph (PI Spectra Pro SP2300). Subsequently, detected with a Streak Camera (Hamamatsu C10910) system with a temporal resolution of 1.4 ps. The data was acquired in photon counting mode using the Streak Camera software (HPDTA), and exported to Origin Pro 2020 for further analysis.

**X-ray photoemission spectroscopy (XPS):** X-ray photoemission spectroscopy (XPS) measurements are performed with a Thermofisher Escalab 250Xi system. All samples were measured under ultrahigh vacuum ($10^{-10}$ mbar). XPS measurements were performed using an XR6 monochromated AlKα source (hv = 1486.6 eV) and a pass energy of 20 eV. Argon cluster ion beam etching experiments were performed using a MAGCIS ion gun using a cluster energy of 4000 eV.

**Author Contributions**
J.I.K. and M.G. contributed equally to this work.

**Supporting Information**

Supporting Information is available from the Online Library or from the authors.

**Acknowledgements**

This publication is based upon work supported by the King Abdullah University of Science and Technology (KAUST) Office of Sponsored Research (OSR) under Award No: OSR-CARF/CCF-3079, OSR-2018-CRG7-3737, OSR-2019-CRG8-4093, OSR-2020-CRG9-4350, and OSR-CRG2018-3783. This project has received funding from the European Research Council (ERC) under the European Union's Horizon 2020 research and innovation programme (ERC Grant

**Notes**

The authors declare no competing financial interest.

**AUTHOR INFORMATION**
Corresponding Authors email:

**References**

(1) Zheng, Y.; Niu, T.; Ran, X.; Qiu, J.; Li, B.; Xia, Y.; Chen, Y.; Huang, W. Unique Characteristics of 2D Ruddlesden-Popper (2DRP) Perovskite for Future Photovoltaic Application. *J. Mater. Chem. A* **2019**, *7* (23), 13860–13872. https://doi.org/10.1039/c9ta03217g.

(2) Kakavelakis, G.; Gedda, M.; Panagiotopoulos, A.; Kymakis, E.; Anthopoulos, T. D.; Petridis, K. Metal Halide Perovskites for High-Energy Radiation Detection. *Adv. Sci.* **2020**, *7* (22), 2002098. https://doi.org/10.1002/ADVS.202002098.

(3) Gedda, M.; Faber, H.; Petridis, K.; Anthopoulos, T. D. Metal Halide Perovskites for High-Energy Radiation Detection. *Adv. Mater. Radiat. Detect.* **2022**, 119–144. https://doi.org/10.1007/978-3-030-76461-6_6.

(4) Ruddlesden, S. N.; Popper, P. New Compounds of the K2NIF4 Type. *Acta Crystallogr.* **1957**, *10* (8), 538–539. https://doi.org/10.1107/S0365110X57001929.

(5) Stoumpos, C. C.; Cao, D. H.; Clark, D. J.; Young, J.; Rondinelli, J. M.; Jang, J. I.; Hupp, J. T.; Kanatzidis, M. G. Ruddlesden–Popper Hybrid Lead Iodide Perovskite 2D Homologous Semiconductors. *Chem. Mater.* **2016**, *28* (8), 2852–2867. https://doi.org/10.1021/ACS.CHEMMATER.6B00847.

(6) Dou, L.; Wong, A. B.; Yu, Y.; Lai, M.; Kornienko, N.; Eaton, S. W.; Fu, A.; Bischak, C. G.; Ma, J.; Ding, T.; et al. Atomically Thin Two-Dimensional Organic-Inorganic Hybrid Perovskites. *Science (80-. ).* **2015**, *349* (6255), 1518–1521. https://doi.org/10.1126/SCIENCE.AAC7660.

(7) Blancon, J.-C.; Stier, A. V.; Tsai, H.; Nie, W.; Stoumpos, C. C.; Traoré, B.; Pedesseau, L.; Kepenekian, M.; Katsutani, F.; Noe, G. T.; et al. Scaling Law for Excitons in 2D Perovskite Quantum Wells. *Nat. Commun. 2018 91* **2018**, *9* (1), 1–10. https://doi.org/10.1038/s41467-018-04659-x.

(8) Proppe, A. H.; Quintero-Bermudez, R.; Tan, H.; Voznyy, O.; Kelley, S. O.; Sargent, E. H. Synthetic Control over Quantum Well Width Distribution and Carrier Migration in Low-Dimensional Perovskite Photovoltaics. *J. Am. Chem. Soc.* **2018**, *140* (8), 2890–2896. https://doi.org/10.1021/jacs.7b12551.

(9) Wang, Z.; Wang, F.; Zhao, B.; Qu, S.; Hayat, T.; Alsaedi, A.; Sui, L.; Yuan, K.; Zhang, J.; Wei, Z.; et al. Efficient Two-Dimensional Tin Halide Perovskite Light-Emitting Diodes via a Spacer Cation Substitution Strategy. *J. Phys. Chem. Lett.* **2020**, *11* (3), 1120–1127. https://doi.org/10.1021/ACS.JPCLETT.9B03565.

(10) Huang, Y.; Li, Y.; Lim, E. L.; Kong, T.; Zhang, Y.; Song, J.; Hagfeldt, A.; Bi, D. Stable Layered 2D Perovskite Solar Cells with an Efficiency of over 19% via Multifunctional Interfacial Engineering. *J. Am. Chem. Soc.* **2021**, *143* (10), 3911–3917. https://doi.org/10.1021/JACS.0C13087.

(11) Liu, P.; Han, N.; Wang, W.; Ran, R.; Zhou, W.; Shao, Z. High-Quality Ruddlesden–Popper Perovskite Film Formation for High-Performance Perovskite Solar Cells. *Adv. Mater.* **2021**, *33* (10), 2002582. https://doi.org/10.1002/ADMA.202002582.

(12) Gao, X.; Zhang, X.; Yin, W.; Wang, H.; Hu, Y.; Zhang, Q.; Shi, Z.; Colvin, V. L.; Yu, W. W.; Zhang, Y. Ruddlesden–Popper Perovskites: Synthesis and Optical Properties for Optoelectronic Applications. *Adv. Sci.* **2019**, *6* (22), 1900941. https://doi.org/10.1002/ADVS.201900941.

(13) Chen, Y.; Sun, Y.; Peng, J.; Tang, J.; Zheng, K.; Liang, Z. 2D Ruddlesden–Popper Perovskites for Optoelectronics. *Adv. Mater.* **2018**, *30* (2), 1703487. https://doi.org/10.1002/ADMA.201703487.

(14) Cao, D. H.; Stoumpos, C. C.; Farha, O. K.; Hupp, J. T.; Kanatzidis, M. G. 2D Homologous Perovskites as Light-Absorbing Materials for Solar Cell Applications. *J. Am. Chem. Soc.* **2015**, *137* (24), 7843–7850. https://doi.org/10.1021/JACS.5B03796.

(15) Ren, M.; Cao, S.; Zhao, J.; Zou, B.; Zeng, R. Advances and Challenges in Two-Dimensional Organic–Inorganic Hybrid Perovskites Toward High-Performance Light-Emitting Diodes. *Nano-Micro Lett. 2021 131* **2021**, *13* (1), 1–36. https://doi.org/10.1007/S40820-021-00685-5.

(16) Yang, X.; Zhang, X.; Deng, J.; Chu, Z.; Jiang, Q.; Meng, J.; Wang, P.; Zhang, L.; Yin, Z.; You, J. Efficient Green Light-Emitting Diodes Based on Quasi-Two-Dimensional Composition and Phase Engineered Perovskite with Surface Passivation. *Nat. Commun. 2018 91* **2018**, *9* (1), 1–8. https://doi.org/10.1038/s41467-018-02978-7.

(17) La-Placa, M.-G.; Longo, G.; Babaei, A.; Martínez-Sarti, L.; Sessolo, M.; Bolink, H. J. Photoluminescence Quantum Yield Exceeding 80% in Low Dimensional Perovskite Thin-Films via Passivation Control. *Chem. Commun.* **2017**, *53* (62), 8707–8710. https://doi.org/10.1039/C7CC04149G.

(18) Gélvez-Rueda, M. C.; Fridriksson, M. B.; Dubey, R. K.; Jager, W. F.; van der Stam, W.; Grozema, F. C. Overcoming the Exciton Binding Energy in Two-Dimensional Perovskite Nanoplatelets by Attachment of Conjugated Organic Chromophores. *Nat. Commun. 2020 111* **2020**, *11* (1), 1–9. https://doi.org/10.1038/s41467-020-15869-7.

(19) Blancon, J.-C.; Stier, A. V; Tsai, H.; Nie, W.; Stoumpos, C. C.; Traoré, B.; Pedesseau, L.; Kepenekian, M.; Katsutani, F.; Noe, G. T.; et al. Scaling Law for Excitons in 2D Perovskite Quantum Wells. https://doi.org/10.1038/s41467-018-04659-x.

(20) Ishihara, T.; Takahashi, J.; Goto, T. Exciton State in Two-Dimensional Perovskite Semiconductor (C10H21NH3)2PbI4. *Solid State Commun.* **1989**, *69* (9), 933–936. https://doi.org/10.1016/0038-1098(89)90935-6.

(21) Takeoka, Y.; Fukasawa, M.; Matsui, T.; Kikuchi, K.; Rikukawa, M.; Sanui, K. Intercalated Formation of Two-Dimensional and Multi-Layered Perovskites in Organic Thin Films. *Chem. Commun.* **2005**, No. 3, 378–380. https://doi.org/10.1039/B413398F.

(22) Stoumpos, C. C.; Cao, D. H.; Clark, D. J.; Young, J.; Rondinelli, J. M.; Jang, J. I.; Hupp, J. T.; Kanatzidis, M. G. Ruddlesden-Popper Hybrid Lead Iodide Perovskite 2D Homologous Semiconductors. *Chem. Mater.* **2016**, *28* (8), 2852–2867. https://doi.org/10.1021/acs.chemmater.6b00847.

(23) Takeoka, Y.; Fukasawa, M.; Matsui, T.; Kikuchi, K.; Rikukawa, M.; Sanui, K. Intercalated Formation of Two-Dimensional and Multi-Layered Perovskites in Organic Thin Films. *Chem. Commun.* **2005**, *3* (3), 378–380. https://doi.org/10.1039/B413398F.

(24) Zhang, Y.; Liu, Y.; Xu, Z.; Ye, H.; Li, Q.; Hu, M.; Yang, Z.; Liu, S. Two-Dimensional (PEA) 2 PbBr 4 Perovskite Single Crystals for a High Performance UV-Detector. *J. Mater. Chem. C* **2019**, *7* (6), 1584–1591. https://doi.org/10.1039/c8tc06129g.

(25) Gedda, M.; Yengel, E.; Faber, H.; Paulus, F.; Kreß, J. A.; Tang, M.-C.; Zhang, S.; Hacker, C. A.; Kumar, P.; Naphade, D. R.; et al. Ruddlesden–Popper-Phase Hybrid Halide Perovskite/C8-BTBTall-Molecule Organic Blend Memory Transistors. *Adv. Mater.* **2021**, *33* (7), 2003137. https://doi.org/10.1002/ADMA.202003137.

(26) Even, J.; Pedesseau, L.; Katan, C. Understanding Quantum Confinement of Charge Carriers in Layered 2D Hybrid Perovskites. *ChemPhysChem* **2014**, *15* (17), 3733–3741. https://doi.org/10.1002/cphc.201402428.

(27) Liang, M.; Lin, W.; Zhao, Q.; Zou, X.; Lan, Z.; Meng, J.; Shi, Q.; Castelli, I. E.; Canton, S. E.; Pullerits, T.; et al. Free Carriers versus Self-Trapped Excitons at Different Facets of Ruddlesden-Popper Two-Dimensional Lead Halide Perovskite Single Crystals. *J. Phys. Chem. Lett.* **2021**, *12* (20), 4965–4971. https://doi.org/10.1021/ACS.JPCLETT.1C01148/SUPPL_FILE/JZ1C01148_SI_005.CIF.

(28) Chen, Z.; Li, Z.; Hopper, T. R.; Bakulin, A. A.; Yip, H. L. Materials, Photophysics and Device Engineering of Perovskite Light-Emitting Diodes. *Reports Prog. Phys.* **2021**, *84*

(4), 046401. https://doi.org/10.1088/1361-6633/ABEFBA.

(29) Chen, X.; Lu, H.; Li, Z.; Zhai, Y.; Ndione, P. F.; Berry, J. J.; Zhu, K.; Yang, Y.; Beard, M. C. Impact of Layer Thickness on the Charge Carrier and Spin Coherence Lifetime in Two-Dimensional Layered Perovskite Single Crystals. *ACS Energy Lett.* **2018**, *3* (9), 2273–2279. https://doi.org/10.1021/ACSENERGYLETT.8B01315/SUPPL_FILE/NZ8B01315_SI_001.PDF.

(30) Price, M. B.; Butkus, J.; Jellicoe, T. C.; Sadhanala, A.; Briane, A.; Halpert, J. E.; Broch, K.; Hodgkiss, J. M.; Friend, R. H.; Deschler, F. Hot-Carrier Cooling and Photoinduced Refractive Index Changes in Organic–Inorganic Lead Halide Perovskites. *Nat. Commun. 2015 61* **2015**, *6* (1), 1–8. https://doi.org/10.1038/ncomms9420.

(31) Yang, X.; Chu, Z.; Meng, J.; Yin, Z.; Zhang, X.; Deng, J.; You, J. Effects of Organic Cations on the Structure and Performance of Quasi-Two-Dimensional Perovskite-Based Light-Emitting Diodes. *J. Phys. Chem. Lett.* **2019**, *10* (11), 2892–2897. https://doi.org/10.1021/acs.jpclett.9b00910.

(32) Ball, J. M.; Petrozza, A. Defects in Perovskite-Halides and Their Effects in Solar Cells. *Nat. Energy 2016 111* **2016**, *1* (11), 1–13. https://doi.org/10.1038/nenergy.2016.149.

(33) Stokes, A. R.; Wilson, A. J. C. A Method of Calculating the Integral Breadths of Debye-Scherrer Lines. *Math. Proc. Cambridge Philos. Soc.* **1942**, *38* (3), 313–322. https://doi.org/10.1017/S0305004100021988.

**Supporting Information**

**Correlation between the photophysics and photodetector performance of quasi-2 dimensional layered (PEA)2PbBr4 perovskites upon bulk defect passivation**

Jafar I. Khan$*[1], Murali Gedda$[1], Mingcong Wang[1], Emre Yengel[1], Joshua A. Kreß[2], Yana Vaynzof[2], Thomas D. Anthopoulos[1], Frédéric Laquai*[1]

[1]King Abdullah University of Science and Technology (KAUST), KAUST Solar Center (KSC), Physical Sciences and Engineering Division (PSE), Thuwal 23955-6900, Kingdom of Saudi Arabia

[2]Integrated Center for Applied Physics and Photonic Materials, Center for Advancing Electronics Dresden (CFAED), Technical University of Dresden, Nöthnitzer Straße 61, 01187 Dresden, Germany

## Contents

## Crystallite size

The crystallite size (*D*) of the perovskite films were computed from full width half maximum intensity (FWHM) by adopting the Debye-Scherrer equation:[33]

$$D = \frac{k\lambda}{\beta \cos\theta} \quad (1)$$

Where *D* is particle size (nm), *k* is a dimensionless factor which is nearly equal to unity (≈0.94), $\lambda$ is the wavelength (0.154 nm, X-ray CuKα radiation), $\beta$ is FWHM (full width at half maximum), and $\theta$ (theta) is diffraction angle.

Table S1: Calculated crystallite sizes of (PEA)$_2$PbBr$_4$ and 25% thin films using the Debye-Scherer equation.

| Material | 2q (°) | FWHM (2q) | Crystallite size (nm) |
|---|---|---|---|
| (PEA)$_2$PbBr$_4$ | 5.28 | 0.1113 | 74 |
| 25% | 5.28 | 0.0676 | 123 |

# Time-resolved photoluminescence

Table S2: TR-PL fitting parameters using a two exponential function fit the kinetics. The parameters $A_1$, $A_2$, $\tau_1$, $\tau_2$ are the respective signal amplitudes and lifetimes, whilst $\tau_{avg}$ is the weighted average lifetime estimated as: $((A_1\tau_1+A_2\tau_2)/(A_1+A_2))$.

| Sample | $A_1$ | $\tau_1$(ps) | $A_2$ | $\tau_2$(ps) | $\tau_{avg}$(ps) |
|---|---|---|---|---|---|
| Pristine | 63 | 28 | 64 | 330 | 181 |
| 10% | 34 | 58 | 60 | 388 | 269 |
| 25% | 26 | 43 | 57 | 477 | 341 |
| 50% | 56 | 52 | 46 | 256 | 144 |

# X-ray photoemission spectroscopy

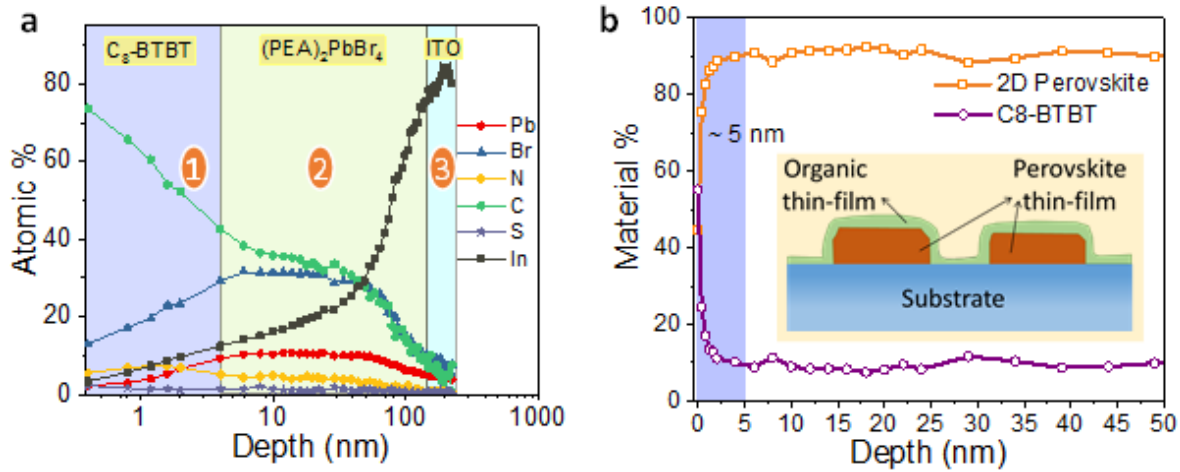

Figure S1: (a) Elemental profiles extracted from X-ray photoemission spectroscopy depth profiling measurements of 2D-RP perovskite: $C_8$-BTBT stack on ITO substrate. Zone-1, 2 and 3 represent the dominant atomic% of $C_8$-BTBT, $(PEA)_2PbBr_4$, and ITO, respectively. In (b) corresponding material % profile is calculated from (a). The numbers in the circles are an identification of the different zones as discussed in the text.

# Transient absorption spectroscopy

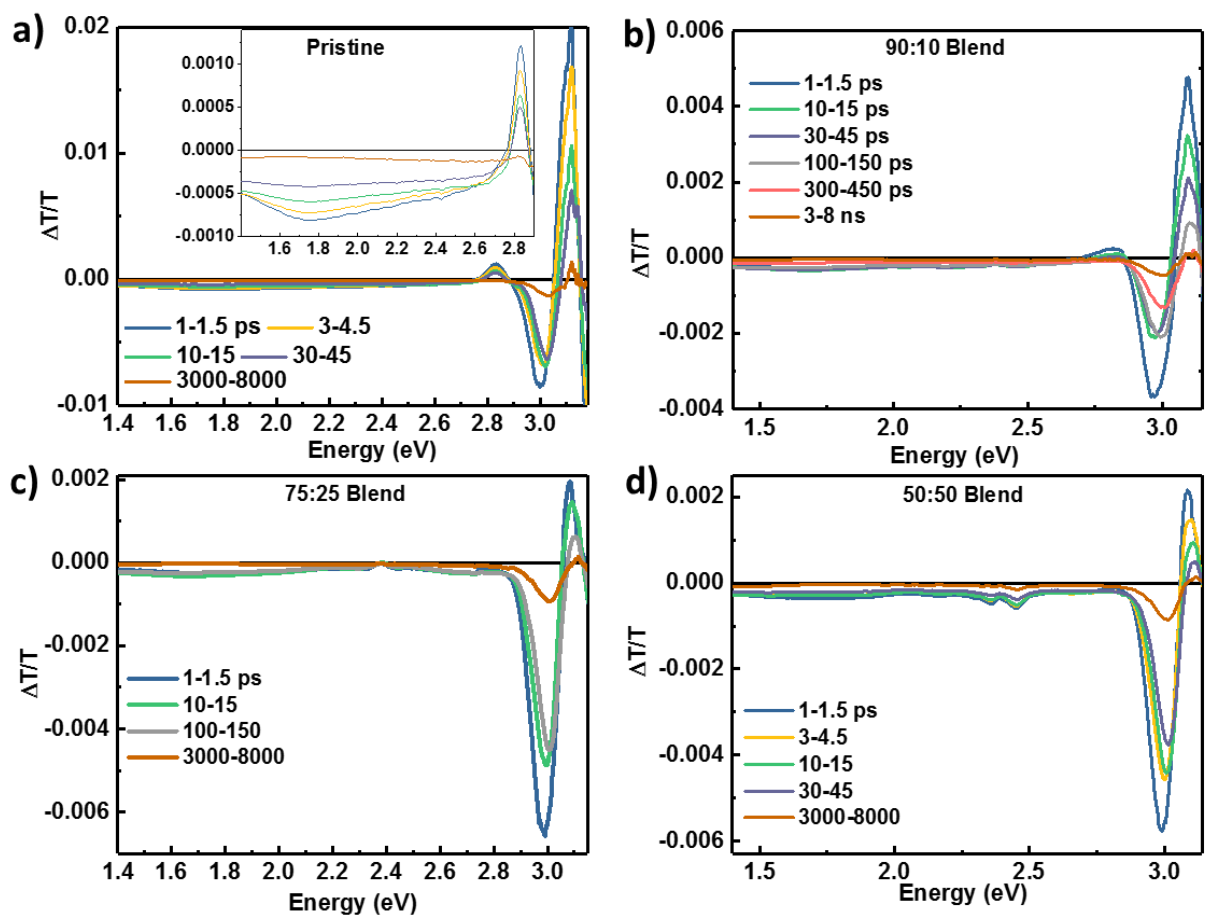

Figure S2: The selected time-integrated TA spectra of the representative systems upon excitation at 380 nm. In a) the pristine 2D perovskite TA spectra with the inset showing the spectral range from 1.4-1.8 eV, b) the 10% blend TA spectra, c) the 25% blend TA spectra, d) the 50% blend TA spectra.